\documentstyle[12pt]{article}
\def\binom#1#2{{#1\choose#2}}
\def\ds{\displaystyle} 
\begin{document}
\newtheorem{theorem}{theorem}
\newtheorem{proposition}{proposition}
\newtheorem{definition}{definition}
\newtheorem{lemma}{lemma}
\newtheorem{notation}{notation}
\begin{flushright}
SISSA-ISAS 53/97/FM\\hep-th/9704138\\
\end{flushright}
\vspace{1cm}
\begin{center}
{\Large {Grassmannian Cohomology Rings and\\
\vspace{.3cm}
Fusion Rings from Algebraic Equations}}\\
\vspace{1cm}
Noureddine Chair\footnote{e-mail:chair@sissa.it}\\
{\it SISSA/ISAS Via Beirut 2, 34014\\Trieste, Italy}\\
\end{center}
\vspace{1.5cm}
\begin{abstract}
\noindent The potential that generates the cohomology ring of the
Grassmannian is given in terms of the elementary symmetric functions
using the Waring formula that computes the power sum of roots of an
algebraic equation in terms of its coefficients. As a consequence, the
fusion potential for $su(N)_K$ is obtained. This potential is the
explicit Chebyshev polynomial in several variables of the first kind.
We also derive the fusion potential for $sp(N)_K$ from a reciprocal
algebraic equation. This potential is identified with another
Chebyshev polynomial in several variables.  We display a connection
between these fusion potentials and generalized Fibonacci and Lucas
numbers.
\end{abstract}
\vfill
\begin{flushleft}
SISSA-ISAS 53/97/FM\\
hep-th/9704138\\
April 1997
\end{flushleft}
\eject
\section{Introduction}
In the work of Gepner \cite{G}, the fusion potential for $su(N)_K$ was
obtained as a perturbation of the Landau-Ginzberg potential
that generates the
cohomology ring of the Grassmannian. This
implies that the fusion ring for $su(N)_K$ and the cohomology ring of
the Grassmannin are connected. The connection of these two rings may
be understood as follows: Lesieur \cite{Le} noticed that the rules of
multiplying Schubert cycles\cite{GH}, which are the generators of the
homology
ring of the Grassmannian, formally coincide with the rules for
multiplying Schur functions \cite{MD}. On the other hand, the
characters of the irreducible representation of $su(N)$ turn out to be
given by the Schur functions \cite{K} with some constraint which is
exactly the perturbation mentioned above. Therefore, we learn that the
product of characters is the same as a product of Schur functions with
this constraint which, in turn, implies the connection between the
cohomology ring for the Grassmannian and fusion ring for $su(N)_K$.

The potential that generates the cohomology ring of the Grassmannian
turns out to be given by a power sum symmetric function in the Chern
roots \cite{V} that we identify with the roots of an algebraic
equation, say of degree $r$, {\it i.e.}, of the form
\begin{equation}
y^r + a_1 y^{r-1} + \cdots + a_{r-1}y + a_r =0\,.
\label{I-one}
\end{equation}
Geometrically, the degree $r$ is the rank of the quotient bundle on
the Grassmannian and the coefficients of the algebraic equation
(elementary symmetric functions) correspond to the Chern classes of
this bundle. With this interpretation in mind, the algebraic equation 
(\ref{I-one}) is nothing but the definition of the Chern classes of a
vector bundle of rank $r$ given by Grothendieck \cite{Gro}, where $y$
is identified with the fundamental class of degree 2 on the 
associated projective bundle.

In this paper, we use the Waring formula to express the power sum
symmetric function in the Chern roots in terms of the elementary
symmetric functions and hence obtain the cohomology potential for the
Grassmannian and the fusion potentials for $su(N)_K$ and
$sp(N)_K$. The algebraic equation from which the $su(N)_K$ fusion
potential is obtained is the
one for which $r=N$ and $a_N=1$, whereas, for $sp(N)_K$, it turns
out to be a reciprocal algebraic equation \cite{B} of degree $2N$,
with the last coefficient equal to one and $a_{2N-i} = a_i$.

In our formulation, the fusion potential written in terms of the
elementary symmetric functions is the explicit generalization of the
Chebyshev polynomial of one variable. Similarly, for the case of
$sp(N)_K$, we obtain another Chebyshev polynomial in several
variables. The one-variable Chebyshev polynomials of the first kind
and second kind are known to be related to the ordinary Lucas numbers
and Fibonacci numbers respectively. In this paper, we find a relation
to the generalized Fibonacci and Lucas numbers for the cases studied
here.

Our paper is organized as follows: Section 2 gives a brief account of
the cohomology ring in order to recall some facts and fix the
notation. Section 3 will be devoted to the cohomology ring potential
and its connection with the fusion ring for $su(N)_K$. The connection
of the later with the generalized Chebyshev polynomial and the numbers
of Fibonacci and Lucas will also be discussed. In section 4, we will
consider the $sp(N)_K$ fusion potential and its connection with the
reciprocal algebraic equation. Here, we will find that the Chebyshev
polynomial associated with $sp(N)_K$ is different from the one for
$su(N)_K$ for $N\neq 1$. In this case, the Fibonacci and Lucas numbers
are of degree $2N$. Our conclusions are outlined in section 5.

\section{The Cohomology Ring}
In this section, we will recall briefly the definition of the
cohomology ring of the grassmannian\cite{BT} and  the coresponding
Landau-Ginsburg formulation \cite{V,G}, in order to fix our notation.
The complex grassmannian manifold here denoted by $G_r(C^n)$ is the
space of r-planes in  $C^n$, its cohomology ring denoted by  
$H^*(G_r(C^n))$ is a
truncated polynomial ring in several variables given by
\begin{equation}
H^*(G_r(C^n))\cong C[x_1,\cdots,x_r,y_1,\cdots,,y_{n-r}]/I\,,
\label{alpha}
\end{equation}
where $x_i=c_i(Q)$ (for $1\leq i\leq r$) are the chern classes of the
quotient  bundle $Q$ of rank $r$, {\it i.e.},
$x_i\in H^{2i}(G_r(C^n))$ and $y_j=c_j(S)$ (for $1\leq j\leq n-r)$
are the chern  classes of the universal bundle $S$ of rank $n-r$.
The ideal $I$ in 
$C[x_1,\cdots,x_r,y_1,\cdots,y_{n-r}]$ is given by 
\begin{equation}
(1+x_1+x_2+\cdots+x_r)(1+y_1+y_2+\cdots+y_{n-r})=1\,,
\label{relxy}
\end{equation}
which is the consequence  of the tautological sequence on $G_r(C^n)$
$$
0\longrightarrow S\longrightarrow V\longrightarrow Q\longrightarrow 0\,,
$$ 
where $ V=G_r(C^n)\times C^n$. By using equation (\ref{relxy}), one
may rewrite $H^*(G_r(C^n))$ as 
\begin{equation}
H^*(G(C^n))\cong C[x_1,\cdots, x_r]/y_j\,,
\label{c}
\end{equation}
where $y_j$ are expressed in terms of $x_i$, and $y_j=0$ for
$n-r+1\leq j\leq n$,  
and $x_0=y_0=1$. The classes $ y_j $ can be written inductively as a
function of $x_1,\cdots,x_r$ via 
\begin{equation}
y_j=-x_1y_{j-1}-\cdots-x_{j-1}y_1-x_j\quad {\rm for}\, j=1,\cdots,
n-r\,. 
\label{e}
\end{equation}
We will give later on an explicit formula for the $ y_j's$ in terms 
of the $x_i's$ without the use of induction .
 
In the Landau-Ginsburg formulation, the  potential that generates the
cohomolgy ring of the grassmannian as explained in \cite{V,G,W}, is
given by 
\begin{equation}
W_{n+1}(x_1,\cdots ,x_r)=\sum_{i=1}^{r} \frac{q_i^{n+1}}{n+1}\,,
\end{equation}
where, $x_i$ and $q_i$ are related by
\begin{equation} 
x_i=\sum_{1\leq l_1 < l_2 \cdots < l_i \leq r} q_{l_1} q_{l_2} \cdots
q_{l_i} \,.
\end{equation}
Usually the description of the cohomology ring is given in terms 
of the $q_i$ variables, however,  in the next section, we will write
down the potential in terms of the $x_i's$, {\it i.e.}, as a solution
to the above system of equations. Note that, as was shown explicitly
in \cite{W},  the cohomology ring of the grassmannian is given by  
\begin{equation}
\frac{\partial W_{n+1}}{\partial x_i}=(-1)^n y_{n+1-i}, \quad {\rm
for} 1 \leq i \leq r\,,
\end{equation}
implying that $d_iW_{n+1}=0$, for $i=1,\cdots, r$.

\section{The Cohomology Ring Potential}
A formula for the Landau-Ginsburg potential $W_{n+1}(x_1,\cdots ,x_r)$
is given in terms of the generators of $H^*(G_r(C^n))$, and when we
consider the potential $W_n(x_1,\cdots ,x_r)$ instead with $n=N+k$,
$r=N$ and $x_N=1$ we obtain the fusion potential of the $ su(N)_K$
\cite{G}. The fusion potential in this formulation is the explicit generalized
Chebyshev polynomial in several variables.  The ordinary Fibonacci and
the Lucas numbers are known to be connected to Chebyshev polynomial of
the second kind and the first kind respectively\cite{R}. Here we will find the
connection between the fusion potential of the $ su(N)_K$ algebra and
the $kth$ order Fibonacci and the Lucas numbers.  The following
formulae for the potential, the classes $ y_j$ in terms of the $x_i$
classes and, in general, the connection between Segre classes of any
vector bundle of rank $n$ in termes of Chern classes are first
proposed, then later proved using the theory of symmetric functions
\cite{M}.

\begin{proposition}
The potential $W_{n+1}(x_1,\cdots ,x_r)$ that generates the cohomology 
ring of the grassmannian $H^*(G_r(C^n))$ in terms of the generators
$x_i=c_i(Q)$ for $1\leq r$ is given by the formula
\begin{eqnarray}
&{ \ds W_{n+1}(x_1,\cdots ,x_r) }=  {\ds \sum_{k_1=0}^{\left
[\frac{n+1}{2}\right]}\cdots
\sum_{k_{r-1}=0}^{\left[\frac{n+1}{r}\right]}
\frac{ (-1)^{k_1+2k_2+\cdots+(r-1)k_{r-1}}}{k_1!\cdots
k_{r-1}!} \times }& \nonumber \\ {}\nonumber\\
& {\ds \frac{\left(n- \sum_{j=1}^{r - 1} j k_j \right)!}
{\left(n+1 - \sum_{j=2}^{r} j k_{j-1} \right)!}
x_1^{n+1-2k_1-\cdots rk_{r-1}} x_2^{k_1}\cdots x_r^{k_{r-1}} }.&
\label{a} 
\end{eqnarray}
The above formula reduces to the fusion potential of $ su(N)_K$
algebra when we consider the potential $W_n(x_1,\cdots ,x_r)$ instead,
with $n=N+k$, $r=N$ and $x_N=1$ which in turn is the explicit
multidimentional analogue of Chebyshev polynomial of the first
kind. Finally the fusion potential and the multidimentional analogue
of the Chebyshev polynomial of the second kind are shown to be related
the $kth$ order Lucas and Fibonacci numbers respectively.
\end{proposition}

To prove the above formula, we use the fundamental theorem on
symmetric functions\cite{MD}, which states that any symmetric function
can be written as a polynomial in the elementary symmetric functions.
The potential for the cohomology ring of the grassmannian,
$H^*(G_r(C^n))$ is generated by
$$ 
W_{n+1}(x_1,\cdots ,x_r)=\sum_{i=1}^{r} \frac{q_i^{n+1}}{n+1},
$$ 
{\it i.e.}, the power sum symmetric functions in the Chern roots
\footnote{$q_i$ are the formal variables satisfying $\sum_{i=0}^r x_i
t^i = \prod_{i=1}^r (1 + q_it)$.},$q_i$.  From \cite{M}, we learn that
there is an explicit formula for the power sum in terms of the
elementary symmetric functions. As a matter of fact, this formula was
given by Waring \cite{B,S} in connection with the theory of algebraic
equations in which he found a general expression for the power sum of
the roots of an algebraic equation of order $r$ in terms of its
coefficients. This formula reads
\begin{equation}
s_n=n\sum \ (-1)^{n+l_1+\cdots+l_r} \frac{(l_1+\cdots+l_r-1)!}
{l_1!\cdots l_r!} x_1^{l_1}\cdots x_r^{l_r}\,, 
\label{b}
\end{equation}
where $s_n$ denotes the power sum, the $x_i's$ are the elementary
symmetric functions, and the summation is taken over all positive
integers or zero such that $l_1+2l_2+\cdots +rl_r =n$.

It is clear from equation (\ref{b}) that we obtain the formula for the
cohomology potential given by equation (\ref{a}): Simply shift $n$ to
$n+1$ in equation (\ref{b}), set $l_1=n+1-2l_2-\cdots -rl_r$ and now
by making the change of variables $l_2=k_1,\cdots,l_r=k_{r-1}$, the
formula is obtained.  To prove that the potential $ W_{n+1}(x_1,\cdots
,x_r)$ generates the cohomology ring, we need the following formula
that relates the Chern classes of the universal bundle, $y_j$, to
the Chern classes of the quotient bundle $x_i$
\begin{eqnarray}
{\ds y_j }&= &  (-1)^j {\ds \sum_{k_1=0}^{\left[\frac{j}{2}\right]}\cdots
\sum_{k_{r-1}=0}^{\left[\frac{j}{r}\right]}
\frac{ (-1)^{k_1+2k_2+\cdots+(r-1)k_{r-1}}}{k_1!\cdots
k_{r-1}!} \times } \nonumber \\ {}\nonumber\\
& & {\ds \frac{\left(j- \sum_{l=1}^{r - 1} l k_l \right)!}
{\left(j - \sum_{l=2}^{r} l k_{l-1} \right)!}
x_1^{j-2k_1-\cdots rk_{r-1}} x_2^{k_1}\cdots x_r^{k_{r-1}} }\,.
\label{k} 
\end{eqnarray} 
Again we can use the theory of symmetric functions to prove the
equation. This time however, we use the relation between the
homogeneous product sum (also called the completely symmetric
functions) and the elementary symmetric functions. The Segre classes,
denoted by $s_j$, of a vector bundle of rank $n$  has an expression 
similar to that for the Chern classes of the universal bundle but with
$r=n$ and with $j$ allowed to take the values $1,2,\cdots,n$.   Now
the proof that $ W_{n+1}(x_1,\cdots ,x_r)$ generates the cohomology
ring follows by differenting this potential with respect to $x_i$,
$1\leq i \leq r$. Thus, we obtain
\begin{eqnarray}
&{\ds \frac{\partial W_{n+1}}{\partial x_i}  }=   (-1)^{i-1} 
{\ds \sum_{k_1=0}^{\left[\frac{n+1-i}{2}\right]}
\cdots
\sum_{k_{r-1}=0}^{\left[\frac{n+1-i}{r}\right]}
\frac{ (-1)^{k_1+2k_2+\cdots+(r-1)k_{r-1}}}{k_1!\cdots
k_{r-1}!} \times }& \nonumber \\ {}\nonumber\\
& {\ds \frac{\left(n+1-i- \sum_{j=1}^{r - 1} j k_j \right)!}
{\left({n+1-i} - \sum_{j=2}^{r} j k_{j-1} \right)!}
x_1^{{n+1-i}-2k_1-\cdots rk_{r-1}} x_2^{k_1}\cdots x_r^{k_{r-1}} }&\,. 
\label{l} 
\end{eqnarray}
>From equation (\ref{k}), we see that $y_{n+1-i}$ is exactly the
expression for $ \frac{\partial W_{n+1}}{\partial x_i}$ up to $(-1)^n$
which is zero for $i=1, \cdots ,r$, by definition of the cohomology
ring. Therefore, $\frac{\partial W_{n+1}}{\partial x_i}= (-1)^n
y_{n+1-i}$ implying that $d_iW=0$ for $i=1,\cdots, r$. This shows the
isomorphism between the usual definition of the cohomology ring of the
grassmannian $H^*(G_r(C^n))$ and the Landau-Ginsburg formulation.
 
Now, we come to the connection between the cohomology potential and
fusion potential of $su(N)_K$ algebra.  We consider the potential
$W_n(x_1,\cdots ,x_r)$ with $n=N+k$, $r=N$ and set $x_N=1$ in
the expression of $ W_n(x_1,\cdots ,x_r)$ to obtain the following
potential
\begin{eqnarray}
&{\ds W_{N+K}(x_1,\cdots ,x_N=1) }=  
{\ds \sum_{k_1=0}^{\left[\frac{N+K}{2}\right]}\cdots
\sum_{k_{N-1}=0}^{\left[\frac{N+K}{N}\right]}
\frac{ (-1)^{k_1+2k_2+\cdots+(N-1)k_{N-1}}}{k_1!\cdots
k_{N-1}!} \times }& \nonumber \\ {}\nonumber\\
& {\ds \frac{\left(N+K-1- \sum_{j=1}^{N - 1} j k_j \right)!}
{\left(N+K - \sum_{j=2}^{N} j k_{j-1} \right)!}
\; x_1^{N+K-2k_1-\cdots -Nk_{N-1}} x_2^{k_1}\cdots x_{N-1}^{k_{N-2}}},&
\label{m} 
\end{eqnarray}
this potential is no longer quasihomogeneous. The quasihomogeneous 
part of this potential is obtained by setting $k_{N-1}=0$. To see that
this potential is the natural analogue of Chebyshev  polynomial of the
first kind in several variables, we specialize the potential to the
case of $su(2)_K$ and find
\begin{equation}
{(2+K) W_{2+K}(x) }= (2+K) { \sum_{l=0}^{\left[\frac{K+2}{2}\right]}
\frac{ (-1)^{l} }{l!}
\times  
\frac{(K+1- l)!}{(K+2 - 2l)!}
x^{K+2-2l}}\,.
\label{mu}
\end{equation}
By setting $n=K+2$, one has
\begin{equation}
{n W_{n}(x) }= n { \sum_{l=0}^{\left[\frac{n}{2}\right]}
\frac{ (-1)^{l} }{l!}
\times  
\frac{(n-1- l)!}{(n - 2l)!}
x^{n-2l}}\,.
\label{n}
\end{equation}
This is exactly the Chebyshev polynomial of the first kind
\cite{R}. In this representaion  the Chebyshev polynomial is monic and
with integer coefficients. 
   
It remains to be seen that the analogue of the Chebyshev polynomial of
the first kind in several variables is the fusion ring of the
$su(N)_K$ algebra.  This is a simple consequence of the relation
between our cohomology potential and the Chern classes of the
universal bundle $S$. By using equation (\ref{l}) in the $su(N)_K$
case, one has
\begin{equation}
y_{N+K-i}=(-1)^{i+1} \frac{\partial W_{N+K-i}}{\partial x_i }\quad
{\rm for}\, 1 \leq i \leq N-1,
\label{o}
\end{equation}
which is the ideal of the fusion ring for $su(N)_K$\cite{G}. Therefore
the fusion ring for $su(N)_K$ is 
$R = C[x_1,\cdots x_{N-1}]/(y_{K+1},y_{K+2},\cdots,y_{K+N-1})$.
In terms of Young tableaux, this is equivalent to setting to zero
all reduced tableaux (no columns with N boxes) for which the first row
has length equal to $K+1$- this is the level truncation. The 
Giambeli-like formula \cite{G} when applied to the completely symmetric
representation, then the fusion ideal for $su(N)_K$ reads
\begin{equation}
\underbrace{[1,\cdots 1]}_{j}
= det\, x_{1+l-s},\quad {\rm for}\,\,
1\leq l,s\leq j,\, K+1 \leq j \leq N+K-1.
\end{equation}
Therefore the completely symmetric function $y_j$ given by (\ref{k})
is the explicit expression for the Giambeli-like formula when restricted
to $[1,\cdots 1]$ (with j entries), where $K+1 \leq j \leq N+K-1$.

>From Gepner \cite{G} we learn that there are two ways to obtain the
fusion potential for $su(N)_K$ algebra. One way is to use the
following expresion
\begin{equation}
W_{N+K}(x_1,\cdots ,x_N=1) = \left.\frac{(-1)^{N+K}}{(N+K)!}
\frac{d^{N+K}}{dt^{N+K}}
\log\left(\sum_{i=0}^{N}(-1)^ix_it^i\right)\right|_{t=0}, 
\end{equation}
with $x_0=x_N=1$. Alternatively, we use the recursion relation 
satisfied by the potential
\begin{equation}
\sum_{i=0}^{N}(-1)^{i}x_i(N+s-i)W_{N+s-i}=0\, .
\end{equation}
Therefore, our expression for the fusion potential is simpler and more
transparent. It gives the integrability of the Chern classes $y_j$
(completely symmetric functions) to a potential as a consequence of
the cohomology of the Grassmannian. Furthermore, from our fusion
potential which is the explicit Chebyshev polynomial of the first kind in
several variables, one can read off directly the $su(N)_K$ fusion
potential for any $N$ and $K$.
 
Before we make the connection between the fusion potential of
$su(N)_K$ algebra and the Fibonacci numbers and Lucas numbers of $kth$
order, we will first give the definition of these numbers. We will
then recall the connection between the ordinary Fibonacci and Lucas
numbers with the Chebyshev polynomial of one variable.
\begin{definition}
The $kth$ order Fibonacci numbers $F_{n+1}$ and Lucas numbers $L_n$ are
defined, respectively, by $F_{n+1}=F_n+F_{n-1}+\cdots +F_{n-k}$,
$L_n=L_{n-1}+L_{n-2}+\cdots L_{n-k}$, with the initial conditions
$F_{-k +1} = \cdots = F_{-1} = 0$ and similarly for the $L_n$'s.
\end{definition}

For $k=2$, these are the definitions of the ordinary Fibonacci and
Lucas numbers which are given by $F_{n+1}=F_n+F_{n-1}$ and
$L_n=L_{n-1}+L_{n-2}$, {\it i.e.}, any number is the sum of the
previous two. The Chebyshev polynomial of the second kind
$U(\frac{x}{2})$ is known to be related to the ordinary Fibonacci
numbers, and The Chebyshev polynomial of the of the first kind
$T(\frac{x}{2})$ is known to be related to the Lucas numbers \cite{R}
via the following specializations:
\begin{equation}
F_{n+1}=\frac{S_n(i)}{i^n}, \; \; \; \quad  n=0,1\cdots, (i^2=-1),
\label{om}
\end{equation}
where,
\begin{equation}
S_n(x)=U\left(\frac{x}{2}\right)=\sum_{k=0}^{\left[\frac{n}{2}\right]}(-1)^{k}
\binom{n-k}{k}x^{n-2k}\, ;
\label{p}
\end{equation}
and
\begin{equation}
L_n=\frac{C_n(i)}{i^n}\quad\quad n=0,1\cdots,
\label{q}
\end{equation}
where,
\begin{equation}
 C_n(x)=2T\left(\frac{x} {2}\right) = 
\sum_{k=0}^{\left[\frac{n}{2}\right]} (-1)^{k}\frac{n}{n-k}\binom{n-k}{k}
x^{n-2k}\,.
\label{r}
\end{equation}
By applying a similar procedure, the analogue of the Chebyshev
polynomial of the second kind in several variables and the fusion
potential reduce to the following two sequences of numbers,
respectively:
\begin{equation}
F_{n+1}= \sum_{l_1=0}^{\left[\frac{n}{2}\right]}
\cdots\sum_{l_{k-1}=0}^{\left[\frac{n}{k}\right]}
\frac{1}{l_1!\cdots l_{k-1}!}
\frac{\left(n- \sum_{j=1}^{k - 1} j l_j \right)!}
{\left(n - \sum_{j=2}^{k} j l_{j-1} \right)!}\,,
\label{fib}
\end{equation}
\begin{equation}
L_n= n\sum_{l_1=0}^{\left[\frac{n}{2}\right]}
\cdots\sum_{l_{k-1}=0}^{\left[\frac{n}{k}\right]}
\frac{1}{l_1!\cdots k_{k-1}!}
\frac{\left(n- 1-\sum_{j=1}^{k - 1} j l_j \right)!}
{\left(n - \sum_{j=2}^{k} j l_{j-1} \right)!}
\label{luke}
\end{equation}
One can see that these numbers are indeed those given in 
the definition above, for example, for $k=2$ they are 
the ordinary Fibonacci and Lucas numbers respectively. For $k=3$  
we have the third order Fibonacci and Lucas numbers and we have
computed the first few of them as given below
\begin{equation}
F_{n+1}^{(3)} = 1,\, 1,\, 2,\, 4,\, 7,\, 13,\, 24,\, 44,\, 81,\,\cdots\,,
\label{sheqf}
\end{equation}
\begin{equation}
L_n^{(3)} = 1,\, 3,\, 7,\, 11,\, 21,\, 39,\, 71,\, 131,\,\cdots\,.
\label{sheql}
\end{equation}

In terms of the level $K$, and for a fixed value of $N$ in the
$su(N)_K$, one can see that the first term in the Fibonacci sequence
will start at $n = N + K +1$, whereas that of the Lucas sequence will
start at $n = N +K$, and $N$ is identified with the order of these two
series.
 
The formula given above corresponding to $kth$ order Fibonacci numbers
is in a full agreement with that obtained by Lascoux \cite{L} in which
he showed by using the theory of symmetric functions that $kth$ order
Fibonacci numbers are given by the following multinomial
\begin{eqnarray}
F_{n+1} & =  & \sum_{I} 
     \left(
       \begin{array}{c}
          \ell (I)  \\
          m_{1}, \cdots m_{k}
       \end{array}
     \right)
\label{multif}
\end{eqnarray}
where the summation is taken over all partitions 
$I = 1^{m_1}2^{m_2}\cdots$ of weight $n = m_1 + 2m_2 + \cdots + km_k$ and
$\ell (I)$ is the length of the partition $m_1 + m_2 + \cdots + m_k$. 

The equivalence of our formula for the kth order Fibonacci numbers
and those given by Lascoux follows by expanding the multinomial
(\ref{multif}), and fixing $m_1$ as $m_1 = n - 2m_2 - \cdots - km_k$ and 
then changing the variables as we did before to obtain the 
cohomology potential.

Although the expression for the $kth$ order Lucas number were not 
given in \cite{L}, we can see however that the equivalent 
formula for these numbers is
\begin{eqnarray}
L_{n} & =  & n \sum_{I} 
     \left(
       \begin{array}{c}
          \ell (I) - 1 \\
          m_{1}, \cdots m_{k}
       \end{array}
     \right).
\label{multil}
\end{eqnarray}

\section{$sp(N)_K$ Fusion Potential and the Reciprocal Algebraic Equation}
We recall from the last section that the Waring formula computes the power
sum of roots of an algebraic equation in terms of it
coefficients. These coefficients are identified with the elementary
symmetric functions, in terms of the Chern roots they are given by 
$x_i= \sum_{1\leq l_1,\dots, l_i\leq r} q_{l_1}q_{l_2}\cdots
q_{l_i}$. The algebraic equation from which one computes the
cohomology ring potential has the form 
\begin{equation}
y^r + a_1 y^{r-1} + \cdots + a_{r-1} y + a_r = 0,
\label{III-one}
\end{equation}
where the coefficients $a_i$ are identified with the elementary
symmetric functions, and the roots of this algebraic equation are 
$q_i$ for $i=1,\cdots, r$.

The fusion potential for the $su(N)_K$ algebra may be obtained from
the following algebraic equation
\begin{equation}
y^N + a_1 y^{N-1} + \cdots + a_{N-1} y + 1 = 0
\label{III-two}
\end{equation}
The last coefficient is set equal to $1$ due to the constraints 
$x_N= q_1 q_2 \dots q_N =1$, which in turn corresponds to the fact
that the determinant of the maximal torus of $SU(N)$ group is the
identity. The diagonal elements of this torus are 
$q_i= e^{i(\theta_i - \theta_{i-1})}$ for $i=1,\dots, N$ with the
convention $\theta_0 = \theta_N = 1$. With this motivation in mind, one
would like to know whether one can write down an algebraic equation
corresponding to groups other than $SU(N)$, in particular, the unitary
symplectic group $Sp(N)$. Having written down such an algebraic
equation, the fusion potential for the $sp(N)_K$ algebra is obtained
using the Wearing formula. This turns out to be true as we will
shortly see.

>From \cite{HW} we learn that any $n\times n$ unitary symplectic
matrix (with $n=2m$) can be diagonalized with diagonal elements of the
form $q_i$ and $q_i^{-1}$ for $i=1,\dots, m$, and with determinant
equal to $1$. Therefore the algebraic equation that we are looking 
for is the one for which both $q_i$ and $q_i^{-1}$ are roots and 
where the last coefficient is equal to one.
Such algebraic equations are called reciprocal
equations of the first class \cite{B}. In our case, this algebraic
equation has the form 
\begin{equation}
y^{2m} + a_1 (y^{2m-1} + y) + a_2 (y^{2m -2} + y^2) + \cdots +
a_m y^m +1 =0
\label{III-three}
\end{equation}
where $a_i = a_{2m - i}$. Note that in this case, the elementary symmetric
functions are functions of both $q_i$ and $q_i^{-1}$ that we denote by
$E_i$. Now, the natural power sum to consider for the reciprocal
algebraic equation has the form
$$ 
W_n(E_1,\dots, E_m) = \frac{1}{n} \sum_{i=1}^m (q_i^n + q_i^{-n})
$$
as both $q_i$ and $q_i^{-1}$ are roots of equation (\ref{III-three}).
This is exactly the form proposed by \cite{MS} .
Therefore, by applying the Waring formula to this
expression, one obtains
\begin{eqnarray}
&{W_n(E_1,\dots , E_m) }=   
 {\ds \sum_{k_1=0}^{\left[\frac{n}{2}\right]}\cdots
  \sum_{k_{2m-1}=0}^{\left[\frac{n}{2m}\right]}
\frac{ (-1)^{k_1+2k_2+\cdots+(2m-1)k_{2m-1}}}{k_1!\cdots
  k_{2m-1}!} \times } &\nonumber \\ {}\nonumber\\
  &  {\ds \frac{\left(n-1 - \sum_{l=1}^{2m - 1} l k_l \right)!}
  {\left(n - \sum_{l=2}^{2m} l k_{l-1} \right)!}
  E_1^{g(n,m)} E_2^{k_1+ k_{2m-1}}
\cdots E_m^{k_{m-1}} }\, ,&
  \label{III-four} 
\end{eqnarray} 
where $g(n,m) = {n-2k_1-\cdots - (2m-2)k_{2m-2} - 2m k_{2m-1}}$.
In obtaining the above equation, we have used the condition 
$l_1+2l_2+\cdots + 2m l_{2m} =n$, and the change of variables 
$l_2=k_1, \cdots, l_{2m}=k_{2m-1}$.

In the following we will briefly recall the classical tensor ring for
$sp(N)$  and the modified fusion ring \cite{MS,C}, namely, the
$sp(N)_K$ algebra and hence write explicitly the fusion potential
for the latter. The classical tensor ring for $sp(N)$ is
the finite ring, ${\sc R} = C[\chi_1,\cdots,\chi_N]/I_C$, where
$\chi_j$ are the characters of the fundamental representation
corresponding to a single column of length $j$. These characters are
related to the elementary symmetric function $E_j$ \cite{FH} by
\begin{equation}
\chi_j = E_j - E_{j-2}
\label{III-five}
\end{equation}
The classical ideal $I_C$ is obtained by using this equation with the
property that $E_j = E_{2N-j}$ and $E_0 = E_{2N} = 1$\footnote
{This is exactly the condition that follows from a
reciprocal algebraic equation of degree $2N$, with the last
coefficient $a_{2N}=1$.}, which 
follows from its generating
function \cite{FH}
\begin{equation}
E(t) = \sum_{j=0}^\infty E_j t^j = \prod_{i=1}^N (1+q_it)(1+q_i^{-1}t).
\label{III-six}
\end{equation}
Therefore, the classical ideal $I_C$ is given by 
\begin{eqnarray}
\chi_{N+1} &=& 0, \nonumber\\
\chi_{N+1} + \chi_N &=& 0, \nonumber\\
&\vdots& \nonumber\\
\chi_j + \chi_{2N+2-j} &=& 0. \nonumber 
\end{eqnarray}
The $sp(N)_K$ fusion ring is obtained by a further modification of the 
classical $sp(N)$ tensor ring in which tableaux with more than $K$
columns are eliminated. This is equivalent to writing the $sp(N)_K$-fusion 
ring as ${\sc R}= C[\chi_1\, , \cdots \, , \, \chi_N] /I_f$, where the ideal 
$I_f$ is given by
\begin{eqnarray}
J_{K+1} &=& 0 \, , \nonumber \\
J_{K+2}+ J_K &=& 0 \, , \nonumber \\
&.& \nonumber \\
&.& \nonumber \\
&.&  \nonumber \\
J_{K+N}+J_{K-N+2} &=& 0 \, . 
\label{truncation}
\end{eqnarray}
$J_j$ represents the character of the single row tableaux of length
$j$ which is a completely symmetric function whose generating function 
is \cite{FH},
\begin{equation}
J(t) = \sum_{j=0}^{\infty} J_j t^j = \prod_{i=1}^{N} 
\frac{1}{(1-q_it)(1-q_i^{-1} t)}\; .
\label{character}
\end{equation}
Since $E(t)J(-t)=1$, the completely symmetric functions can be written in 
terms of the elementary symmetric functions as will be 
given explicitly below.

The truncation given by eq.(\ref{truncation}) can be written as the
ideal generated by setting to zero the derivative of the potential 
$W_n$ (\ref{III-four}), for certain values of $n$ and $m$. This
means that the fusion ring for  $sp(N)_K$ can be written as ${\sc R}= 
C[\chi_1\, , \cdots \, , \, \chi_N] /d W_n$ .
To see this we differentiate the potential $W_n$ with respect
to $E_i$ finding,
\begin{equation}
\frac{\partial W_n}{\partial E_i} = \left\{ \begin{array}{ll}
(-)^{i+1}(J_{n-i}+J_{n+i-2m}) \;& \mbox{for} \; 1\leq i \leq n-1 \; 
\nonumber \\
\\
(-)^{i+1}J_{n-i}, \;& \mbox{for} \; i=m\; ,  
\end{array} \right.
\label{critical}
\end{equation}
where $J_j$ is given explicitly by
\begin{eqnarray}
&{ \ds J_j }=   (-1)^j {\ds \sum_{k_1=0}^{\left[\frac{j}{2}\right]}\cdots
  \sum_{k_{2m-1}=0}^{\left[\frac{j}{2m}\right]}
\frac{ (-1)^{k_1+2k_2+\cdots+(2m-1)k_{2m-1}}}{k_1!\cdots
  k_{2m-1}!} \times } &\nonumber \\ {}\nonumber\\
  &  {\ds \frac{\left(j- \sum_{l=1}^{2m - 1} l k_l \right)!}
  {\left(j - \sum_{l=2}^{2m} l k_{l-1} \right)!}
  E_1^{j-2k_1-\cdots - (2m-2)k_{2m-2} - 2m k_{2m-1}} E_2^{k_1+ k_{2m-1}}
\cdots E_m^{k_{m-1}} }\, .&
  \label{complete} 
\end{eqnarray} 
>From eq..(\ref{critical}), we see that the critical points of the potential
$W_n$, $\frac{\partial W_n}{\partial E_i}=0$ do indeed correspond to the 
fusion ideal $I_f$ provided $n=N+K+1$ and $m=N$.

The fusion potential for $sp(N)_K$ algebra is obtained by using the
relation $\chi_j = E_j - E_{j-2}$. 
Setting $\chi_1=x$ and $\chi_2=y$, the fusion potentials
for $sp(1)_K$ and $sp(2)_K$ are

 \begin{equation}
{(2+K) W_{2+K}(x) }= (2+K) { \sum_{l=0}^{\left[\frac{K+2}{2}\right]}
  \frac{ (-1)^{l} }{l!}
    \frac{(K+1- l)!}{(K+2 - 2l)!}
  x^{K+2-2l}},
  \label{eleven}
  \end{equation}
and
\begin{eqnarray}
&{(3+K)W_{3+K} (x)= (3+K)  \sum_{k_1=0}^{\left[\frac{K+3}{2}\right]}
 \sum_{k_2=0}^{\left[\frac{K+3}{3}\right]}
 \sum_{k_3=0}^{\left[\frac{K+3}{4}\right]}
  \frac{ (-1)^{k_1+2 k_2+3 k_3} }{k_1! k_2! k_3!}\times}
&\nonumber \\ {}\nonumber\\
  & { \frac{(2 + K - k_1 - 2k_2 - 3k_3)!}
{(3 + K - 2k_1 - 3k_2 - 4k_3)!}
  x^{3 + K - 2k_1 - 2k_2 - 4k_3}(1+ y)^{k_1}.}&
  \label{twelve}
  \end{eqnarray}

>From equation (\ref{eleven}) we see that this is the Chebyshev
polynomial of the first kind for $su(2)_K$ as it should be, since
$sp(1)=su(2)$. For levels $K=1$ and $K=2$, equation (\ref{twelve})
gives the following potentials $4W_4 = x^4 -4 x^2 y + 2y^2 + 4y -2$
and $5W_5 = x^5 - 5x^3y + 5xy^2 + 5xy - 5x $. These were the potentials
obtained in \cite{GSh} using the recursion relations.

The generalized Chebyshev polynomial in several variables for
$sp(N)_K$ is obtained from the power sum $W_n$ given in equation
(\ref{III-four}) with $n=N+K+1$ and $m=N$,
\begin{eqnarray}
&{W(E_1,\cdots, E_N)}= 
{\ds \sum_{k_1=0}^{\left[\frac{N + K +1}{2}\right]}\cdots
  \sum_{k_{2N-1}=0}^{\left[\frac{N + K +1}{2N}\right]}
\frac{ (-1)^{k_1+2k_2+\cdots+(2N-1)k_{2N-1}}}{k_1!\cdots
  k_{2N-1}!} \times } &\nonumber \\ {}\nonumber\\
  &  {\ds \frac{\left(N + K - \sum_{l=1}^{2N - 1} l k_l \right)!}
  {\left(N + K +1 - \sum_{l=2}^{2N} l k_{l-1} \right)!}
  E_1^{f(N,K)}
E_2^{k_1+ k_{2N-1}} \cdots E_N^{k_{N-1}} }\, ,&
\label{thirteen}
\end{eqnarray}
where $f(N,K) = {N + K + 1 -2k_1- \cdots - (2N-2)k_{2N-2} - 2N k_{2N-1}}$.
This is different from the generalized Chebyshev polynomial in several
variables for $su(N)_K$ for $N\neq 1$. The above results can be stated
by the following proposition,
\begin{proposition}
The fusion potential for the $sp(N)_K$ algebra is
obtained from the power sum of the roots of a reciprocal algebraic
equation of degree $2N$, the last coefficient of which is $1$. The
associated generalized Chebyshev polynomial of the first kind in terms
of the elementary symmetric functions is given by the equation
(\ref{thirteen}).
\end{proposition}

>From the expression for the Chebyshev polynomial in several variables
(\ref{thirteen}), one notes that the Lucas numbers are
of order $2N$. Therefore these sequences are the same as those associated 
with $su(N)_K$ with N even. The difference however, is that in the latter
the sequence starts at $n = 2N + K$, whereas that associated with 
$sp(N)_K$ starts at $n = N + K + 1$. The point that is interesting 
to note is that the Fibonacci numbers associated to $sp(N)_K$ 
are combinations of two Fibonacci numbers. One can see this 
from (\ref{critical}) by considering the analogue of the Chebyshev polynomial 
that is associated with $sp(N)_K$ of the second kind. These numbers follow
from $\partial W_{N + K + 1}/\partial E_1 = J_{K+N} + J_{K-N+2}, N\neq 1$, 
to give the following Fibonacci type sequences of order $2N$. 
\begin{equation}
\tilde{F}_{N+K+2} = F_{N+K+1} + F_{K-N+3},
\label{nf1}
\end{equation}
where,

  \begin{equation}
 F_j= \sum_{k_1=0}^{\left[\frac{j}{2}\right]}
\cdots\sum_{k_{2N-1}=0}^{\left[\frac{j}{2N}\right]}
   \frac{1}{k_1!\cdots k_{2N-1}!}
  \frac{\left(j-\sum_{l=1}^{2N - 1} l k_l \right)!}
  {\left(j - \sum_{l=2}^{2N} l k_{l-1} \right)!.}
  \label{nf2}
 \end{equation}

As an example we have computed the Fibonacci numbers associated with
$sp(2)_K$ using equation \ref{nf1}. The first few numbers are given 
below,
\begin{equation}
3,\, 5,\, 10,\, 19,\, 37,\, 71,\, 137,\, \cdots
\end{equation}
where the first term in this sequence corresponds to formally
to $K=0$. We see that this sequence is a fourth order sequence
and is different from the one obtained for $su(4)_K$.

\section{Conclusions}

In this paper we have seen that the cohomology potential that generates the
cohomology ring of the Grassmannian $G_r(C^n)$, the fusion potentials for
$su(N)_K$ and that for $sp(N)_K$ are obtained from suitable algebraic 
equations using the Waring formula (\ref{b}) that computes the power
sum of roots in terms of the elementary symmetric functions. The roots of 
these algebraic equations are in one to one correspondence with the 
elements of the 
diagonalized form of the unitary matrix groups $U(N)$, $SU(N)$ and $Sp(N)$.

In this algebraic formulation we see clearly that the isomorphism of Lie 
algebras should be translated into the identification of the corresponding 
algebraic equations. For example $su(2)$ and $sp(1)$ have the same algebraic 
equations which follow from eq.(\ref{III-two}) 
and eq.(\ref{III-three}). As 
$sp(2) = so(5)$ the fusion potential for $so(5)_K$ should be
given by eq. (\ref{twelve}) and hence the corresponding algebraic equation is
\begin{equation}
y^4+\chi_1(y^3+y)+(\chi_2+1)y^2 +1 =0 \; .
\end{equation}
Therefore algebraic equations could be used for classifying fusion rings.

In this paper we also obtained an explicit connection 
between the fusion potentials 
and the Chebyshev polynomial in several variables 
for $su(N)_K$, which would be 
difficult to see in the formulation of \cite{G} and \cite{GSh}. These 
polynomials were shown to be related to the Fibonacci and the Lucas numbers.
In the case of $sp(N)_K$ the Fibonacci numbers are of order $2N$ and appear to
be new as they are different from those of $su(N)_K$ of the same 
order, however the Lucas numbers in both cases 
have the same order and belong to the same sequence.

We will see  in our forthcoming paper 
\cite{Chair2-themission} that the ordinary Fibonacci numbers 
arise as intersections numbers on the moduli space of the 
holomorphic map from the sphere $CP^1$ into the Grassmannian $G_2(C^5)$. 

\vspace{1.5cm}
{\centerline{\bf Acknowledgements}}
I would like to thank L. P. Colatto, S. F. Hassan and M. O'Loughlin
for their help during this work and for critical reading of the
manuscript. I would also like to thank M. S. Narasimhan and C. Reina
for discussions and the ICTP and SISSA for support and hospitality.

\end{document}